\documentclass{jfm}

\usepackage{newtxtext}
\usepackage{newtxmath}
\usepackage{natbib}

\usepackage{graphicx}
\usepackage{dcolumn}
\usepackage{bm}
\usepackage{hyperref}
\usepackage{todonotes}
\usepackage{float}
\usepackage{mathtools}
\usepackage{csquotes}
\usepackage{cleveref}
\usepackage{braket}
\usepackage{subcaption}
\usepackage{epstopdf, epsfig}

\newcommand{\beq}{\begin{equation}}
\newcommand{\eeq}{\end{equation}}

\newcommand{\emphasize}[1]{\textit{#1}}
\newcommand{\R}{Re}
\renewcommand{\vec}{\bm}

\newcommand{\review}[1]{{\textcolor{black}{#1}}}

\hypersetup{
    colorlinks = true,
    urlcolor   = blue,
    citecolor  = black,
}

\newcommand{\RomanNumeralCaps}[1]
\linenumbers

\title{Interpreted machine learning in fluid dynamics: Explaining relaminarisation events in wall-bounded shear flows}

\author{Martin Lellep\aff{1}
	\corresp{\email{martin.lellep@ed.ac.uk}},
	Jonathan Prexl\aff{2},
	Bruno Eckhardt\aff{3}
	\and Moritz Linkmann\aff{4}}

\affiliation{
	\aff{1}SUPA, School of Physics and Astronomy, The University of Edinburgh, James Clerk Maxwell
	Building, Peter Guthrie Tait Road, Edinburgh EH9 3FD, UK
	\aff{2}Department of Civil, Geo and Environmental Engineering, Technical University of Munich, D-80333 Munich, Germany
	\aff{3}Physics Department, Philipps-University of Marburg, D-35032 Marburg, Germany
	\aff{4}School of Mathematics and Maxwell Institute for Mathematical Sciences, University of Edinburgh, Edinburgh, EH9 3FD, United Kingdom}

\begin{document}

\maketitle

\begin{abstract}
	Machine Learning (ML) is becoming increasingly popular in fluid dynamics. 
	Powerful ML algorithms such as neural networks or ensemble
	methods are notoriously difficult to interpret. 
        Here, we introduce the novel
	Shapley Additive Explanations (SHAP) algorithm (Lundberg \& Lee,
        2017), a game-theoretic approach that explains the output of a given ML model, in the fluid dynamics context.  
	We give a proof of concept concerning SHAP as an explainable AI method 
	providing useful and human-interpretable insight for fluid dynamics.
	To show that the feature importance ranking provided by SHAP can be interpreted physically,  
	we first consider data from an established low-dimensional model based on the self-sustaining process (SSP) in 
	wall-bounded shear flows, where 
	each data feature 
	has a clear physical and dynamical interpretation in
	terms of known representative features of the near-wall dynamics, i.e.
	streamwise vortices, streaks and linear streak instabilities. 
	\review{
	SHAP determines consistently
	that only 
	the laminar profile, the streamwise vortex, and a specific streak instability
	play a major role in the prediction.}
	We demonstrate that the method can be applied to larger fluid dynamics datasets 
	by a SHAP evaluation on plane Couette flow in a minimal flow unit focussing on the relevance 
	of streaks \review{and their} instabilities for the 
	prediction of relaminarisation events. \review{Here, we find that the prediction is based 
	on proxies for streak modulations corresponding to linear streak instabilities within the SSP.
        That is, the SHAP analysis suggests that the break-up of the self-sustaining cycle is connected with 
	a suppression of streak instabilities.}


\end{abstract}


\section{\label{sec:introduction}Introduction}

Recent successes in the application of artificial intelligence (AI) methods to fluid dynamics cover a
wide range of topics.  These include model building such as a data-driven
identification of suitable Reynolds-averages Navier-Stokes models
\citep{Duraisamy2019, rosofsky2020artificial}, subgrid-scale parametrisations
\citep{xie2020spatially,rosofsky2020artificial}, state estimation by neural
networks based on reduced-order models \citep{Nair2020}, data assimilation for
rotating turbulence \citep{buzzicotti2020reconstruction} through Generative
Adversarial Networks \citep{goodfellow2014generative}, dynamical and statistical prediction
tasks \citep{srinivasan2019predictions,lellep2020using,
boulle2020classification,pandey2020reservoir,Sreenivasan2020}, or pattern extraction
in thermal convection \citep{schneide2018lagrangian,fonda2019pattern}.
Open questions remain as to how AI can be used to
increase our knowledge of the physics of a turbulent flow, which in turn
requires knowledge as to what data features a given machine learning (ML)
method bases its decisions upon. This is related to the question of {\em
representativeness} versus {\em significance} introduced and discussed by
\citet{jimenez2018machine} in the context of two-dimensional homogeneous
turbulence and motivates the application of explainable AI.

Lately, advances in model agnostic explanation techniques have been made by
\citet{lundberg2017unified} in the form of the introduction of
\emphasize{SHapley Additive exPlanations} (SHAP) values. These techniques have
proven themselves useful in a wide range of applications, such as decreasing the
risk of hypoxaemia during surgery \citep{lundberg2018explainable} by indicating
the risk factors on a per-case basis. Subsequently, these methods have been
adapted and optimised for tree ensemble methods \citep{lundberg2018consistent}. Here, we use boosted trees as well as deep neural networks in conjunction with SHAP
values to provide a first conceptual step towards a machine-assisted
understanding of relaminarisation events in wall-bounded shear flows.

Relaminarisation describes the collapse of turbulent transients onto a linearly
stable laminar flow profile. It is intrinsically connected with the transition
to sustained turbulence in wall-bounded shear flows. Localised turbulent
patches such as puffs in pipe flow 
either relaminarise or split in two \citep{wygnanski1973puffsplitting,Nishi2008puffsplitting,Avila2011}.
Transient turbulence is explained in dynamical systems terms through a boundary
crisis between a turbulent attractor and a lower branch of certain exact
solutions of the Navier-Stokes equations \citep{Kawahara2001,Kreilos2012,Lustro2019}.
In consequence, the boundary of the basin of attraction
of the laminar fixed point becomes fractal, and the turbulent attractor transforms into a chaotic saddle.
Relaminarisation events correspond to state-space trajectories originating within this complex basin of
attraction of the laminar state, eventually leaving the chaotic saddle in favour of the laminar fixed point.
For an ensemble of state-space trajectores, the hallmark of escape from a chaotic saddle - a memoryless process - is an
exponential sojourn time distribution
$P(t) \propto \exp{(t/\tau)} \ ,$
with $P(t)$ denoting the probability of residing within the strange saddle
after time $t$ and $\tau$ the characteristic time scale of the escape
\citep{ott2002chaos}.
Exponentially distributed sojourn times, or turbulent lifetimes, are a salient feature of
wall-bounded turbulence close to onset, for instance in pipe flow
\citep{hof2006finite,Eckhardt2007,Hof2008,Avila2010,avila2011onset} or plane
Couette flow
\citep{Schmiegel97,bottin1998discontinuous,Eckhardt2007,schneider2010transient,Shi2013},
and they occur in box turbulence with periodic boundary conditions provided the
forcing allows relaminarisation \citep{Linkmann2015}. The associated time scale
$\tau$ usually increases super-exponentially with Reynolds number
\citep{Eckhardt2008,Hof2008,avila2011onset,Linkmann2015}. 
The puff splitting process also has a characteristic Reynolds-number-dependent time scale,
and the transition to sustained and eventually space-filling turbulence occurs when the puff splitting time scale
exceeds the relaminarisation time scale \citep{avila2011onset}. In the language of
critical phenomena, the subcritical transition to turbulence
belongs to the Directed Percolation universality class
\citep{Pomeau86,Lemoult2016}.

In order to facilitate the physical interpretation
and to save computational effort, in this first step we consider 
a nine-dimensional shear flow model \citep{Moehlis2004} that reproduces the aformentioned
turbulence lifetime distribution \citep{Moehlis2004} of a wall-bounded parallel shear flow. 
Subsequently, and in order to demonstrate that the method can be upscaled to larger
datasets relevant to fluid dynamics applications, we provide an example, where the 
same classification task is carried out on data obtained by Direct Numerical Simulation (DNS) of 
plane Couette flow in a minimal flow unit. Here, we focus on the structure of high- and low-speed 
streaks characteristic of near-wall turbulence.

The low-dimensional model is obtained from the Navier-Stokes
equations by Galerkin truncation and the basis functions are chosen to
incorporate the self-sustaining process (SSP) \citep{waleffe1997self}, which
describes the basic nonlinear near-wall dynamics of wall-bounded parallel shear
flows close to the onset of turbulence.  According to the SSP, a streak is
generated by advection of the laminar flow by a streamwise vortex, this streak
is linearly unstable to spanwise and wall-normal perturbations, which couple to
re-generate the streamwise vortex and the process starts anew.  The
nine-dimensional model assigns suitably constructed basis functions to the
laminar profile, the streamwise vortex, the streak and its instabilities, and
includes a few more degrees of freedom to allow for mode couplings.
Each basis function, that is, each feature for the subsequent ML steps, has a clear physical
interpretation.  Hence the model lends itself well for a first application of
explainable AI methods to determine which flow features are significant for
the prediction of relaminarisation events.

The nine-mode model by \citet{Moehlis2004} and similar low-dimensional models have been considered in
a number of contributions addressing fundamental questions in the dynamics of
parallel shear flows. Variants of the nine-mode model have been used, for
instance, to introduce the concept of the edge of chaos to fluid dynamics and
its connection with relaminarisation events \citep{skufca2006edge}, to
understand drag reduction in viscoelastic fluids \citep{Roy2006}, or to develop
data-driven approaches to identify extreme fluctuations in turbulent flows
\citep{Schmid2018}.  In the context of AI, \citet{srinivasan2019predictions}
used different types of neural networks (NNs) to predict the turbulent
dynamics of the nine-dimensional model. There, the focus was on the ability of
NNs to reproduce shear flow dynamics and statistics with a view
towards the development of machine-assisted subgrid-scale models.
Good predictions of the mean streamwise velocity and Reynolds stresses
were also obtained with Echo State Networks (ESNs)
\citep{Sreenivasan2020}.
\citet{DoanPolif19} used physics-informed ESNs,
where the equations of motion are incorporated as an additional term
in the loss function, for dynamical prediction of chaotic bursts related to
relaminarisation attempts.

The key contribution in our work is to identify the significant features within a
data-driven prediction of relaminarisation events, that is,
the features a classifier needs to see in order to perform well.
For the NMM,
apart from the laminar profile, we find that SHAP identifies some of the main
constituents of the self-sustaining process, the streamwise vortex and a single
sinusoidal streak instability, as important for the prediction of
relaminarisation events.  Other features, such as the streak mode or certain
streak instabilities,  which are certainly of relevance for the dynamics, are
not identified.  These strongly correlate with the features that have been
identified as important for the classification, hence they carry little
additional information for the classifier. There is no {\em a-priori}
reason for choosing, say, the streamwise vortex instead of the streak as a
feature relevant for the prediction. In fact, if predictions are run using only
subsets consisting of featured that have not been identified as important but
correlate with important features, the prediction accuracy drops significantly.
Finally, the information provided by SHAP is discussed in conjunction
with the model equations to provide physical insights into the inner workings 
of the SSP within the remit of the nine-mode model. 
For the DNS data, SHAP values indicate that the classifier bases its 
decisions on regions in the flow that can be associated with streak instabilities.
This suggests SHAP as a method to inform the practitioner as to which flow
features carry information relevant to the prediction of relaminarisation
events, information that cannot be extracted by established means.

The remainder of this article is organised as follows. We begin with an
introduction of the nine-mode-model, its mathematical structure and dynamical
phenomenology in sec.~\ref{sec:fluid_dynamical_system}. Subsequently,
sec.~\ref{sec:ML_and_SHAP} summarises the technical details of the
machine-learning approach, that is, boosted trees for the classification and
SHAP values for the interpretation.  The results of the main investigation are
presented in sec.~\ref{sec:results}. First, we summarise the prediction of
relaminarisation events.  Second, the most important features, here the physically 
interpretable basis functions of the aforementioned nine-mode model (NMM),
are identified by ranking according to the mean absolute SHAP values for a number
of prediction time horizons. Short prediction times, where the nonlinear
dynamics is already substantially weakened, serve as validation cases.  As
expected, the laminar mode is the only relevant feature in the prediction in
such cases. For longer prediction times the laminar mode remains important, and
the modes corresponding to the streamwise vortex and the sinusoidal streak
instability become relevant.  Therein, sec.~\ref{sec:interpretation} contains a
critical discussion and interpretation of the results described in the previous
sections. Here, we connect the significant features identified by SHAP to
important human-observed characteristics of wall-bounded shear flows such as
streaks and streamwise vortices in the self-sustaining process.  
Section \ref{sec:SHAP_values_of_DNS} provides an example SHAP calculation on DNS data of 
plane Couette flow in a minimal flow unit. 
%
We summarise
our results and provide suggestions for further research in
sec.~\ref{sec:conclusions} with a view towards the application and extension
of the methods presented here to higher-dimensional data obtained from experiments 
or high-resolution numerical simulations.

\section{\label{sec:fluid_dynamical_system}The nine-mode model}

We begin with a brief description of the nine-mode model \citep{Moehlis2004}
and its main features. The model is obtained by Galerkin truncation of a
variation of plane Couette flow with free-slip boundary conditions at the
confining walls,
the sinusoidal shear flow.  Sinusoidal shear flows show qualitatively similar
behavior compared with canonical shear flows such as pipe and plane Couette
flow, in the sense that (i) the dynamics is goverened by the self-sustaining
process \citep{waleffe1997self}, and (ii) the laminar profile is linearly
stable for all Reynolds numbers \citep{drazin2004hydrodynamic}. Most
importantly, the sinusoidal shear flow we use subcritically transitions to
turbulence and shows relaminarisation events, it is thus a prototypical
example of a wall-bounded shear flow.

More precisely, we consider an incompressible flow of a Newtonian fluid between
two -  in principle -  infinitely extended parallel plates a distance $d$ apart, with
free-slip boundary conditions in the wall-normal $x_2$-direction.  Periodic
boundary conditions in the homogeneous streamwise ($x_1$-) and spanwise
($x_3$)-directions model the infinite extent of the plates. The sinusoidal
shear flow is thus described by the incompressible Navier-Stokes equations in a rectangular
domain $\Omega = [0,L_1] \times [-d/2,d/2] \times [0,L_3]$. These read in non-dimensionalised form
\begin{align}
	\label{eq:momentum}
		\frac{\partial \vec{u}}{\partial t} + (\vec{u}\cdot\vec{\nabla})\vec{u} & = -\vec{\nabla} p + \frac{1}{\R} \vec{\Delta} \vec{u} + \frac{\sqrt{2}\pi^2}{4\R}\sin(\pi x_2 / 2 )\vec{\hat{e}}_{x_1}  \ ,\\
	\label{eq:incomp}
		\vec{\nabla}\cdot\vec{u} & = 0\ ,
\end{align}
where $\vec{u}(\vec{x}=(x_1, x_2, x_3))=(u_1, u_2, u_3)$ is the fluid velocity, $p$ is the pressure divided by the density and $\R=U_0 d/(2 \nu)$ the
Reynolds number based on the kinematic viscosity $\nu$, the velocity of the
laminar flow $U_0$ and the distance $d$ between the confining plates, and
$\vec{\hat{e}}_{x_1}$ the unit vector in the streamwise direction.  The last
term on the right-hand side of eq.~\eqref{eq:momentum} corresponds to an
external volume force, which is required to maintain the flow owing to the
free-slip boundary conditions.  It sustains the laminar profile $\vec{U}(x_2) =
\sqrt{2}\sin(\pi x_2 / 2)\vec{\hat{e}}_{x_1}$ and determines thereby the velocity
scale $U_0$, which is given by $\vec{U}(x_2)$ evaluated at a distance $x_2 = d/4$
from the top plate. The non-dimensionalisation with respect to $U_0$ and $d/2$ results in
time being given in units of $d/(2U_0)$.

The NMM of \citet{Moehlis2004} is a low-dimensional representation of
the sinusoidal shear flow obtained by Galerkin projection onto a subspace spanned by
nine specifically chosen orthonormal basis functions
$\vec{u}_i(\vec{x})$ for $i = 1, \hdots, 9$ with $\langle \vec{u}_i(\vec{x}), \vec{u}_j(\vec{x})
\rangle = \delta_{ij}$, where $\langle \cdot \rangle $ denotes the $L_2$-inner product on $\Omega$.
The nine-mode model extends previous models by Waleffe with 4 and 8 modes
\citep{waleffe1995transition,waleffe1997self} based on the SSP.
Each mode has a clear interpretation,
\begin{align*}
	\vec{u}_1(\vec{x}) = \vec{U}(x_2)                    & \text{ - the laminar profile}, \\
	\vec{u}_2(\vec{x})                                 & \text{ - the streak}, \\
	\vec{u}_3(\vec{x})                                 & \text{ - the downstream vortex}, \\
	\vec{u}_4(\vec{x}) \text{ and } \vec{u}_5(\vec{x}) & \text{ - streak instabilities: spanwise flows}, \\
	\vec{u}_6(\vec{x}) \text{ and } \vec{u}_7(\vec{x}) & \text{ - streak instabilities: normal vortex modes}, \\
	\vec{u}_8(\vec{x})                                 & \text{ - a three-dimensional interaction mode}, \\
	\vec{u}_9(\vec{x})                                 & \text{ - a model for the modification to the laminar profile by Reynolds stresses}.
\end{align*}
The basis functions, or modes, are divergence free and satisfy the aforementioned boundary conditions.
We refer to eqs.~(7) to (16) of \citet{Moehlis2004} for the explicit mathematical expressions.

The Galerkin projection results in the following expansion
\begin{equation}
	\vec{u}(\vec{x}, t) = \sum_{i=1}^{9} a_i(t) \vec{u}_i(\vec{x}),
	\label{eq:nmm_expansion}
\end{equation}
for the velocity field with nine corresponding time-dependent coefficients
$a_i(t)$. Equation \eqref{eq:momentum} then gives rise
to a system of nine ordinary differential equations for $a_1(t), \hdots, a_9(t)$
- the NMM - given by eqs.~(21) to (29) of \citet{Moehlis2004}.
Despite its simplicity, the dynamics of the NMM resembles that of wall-bounded
shear flows close to the onset of turbulence which transition subcritically.
First, it is based on the near-wall cycle, the SSP, by construction.
Secondly, its transient chaotic dynamics collapses onto the laminar fixed point
with exponentially distributed lifetimes \citep[Fig.7]{Moehlis2004}, that is,
it shows relaminarisation events with qualitatively similar statistics as
wall-bounded parallel shear flows.
Hence, the model is suitable for a study concerned with the prediction of
relaminarisation events of turbulent shear flows.

\begin{figure}
	\centerline{\includegraphics[scale=1]{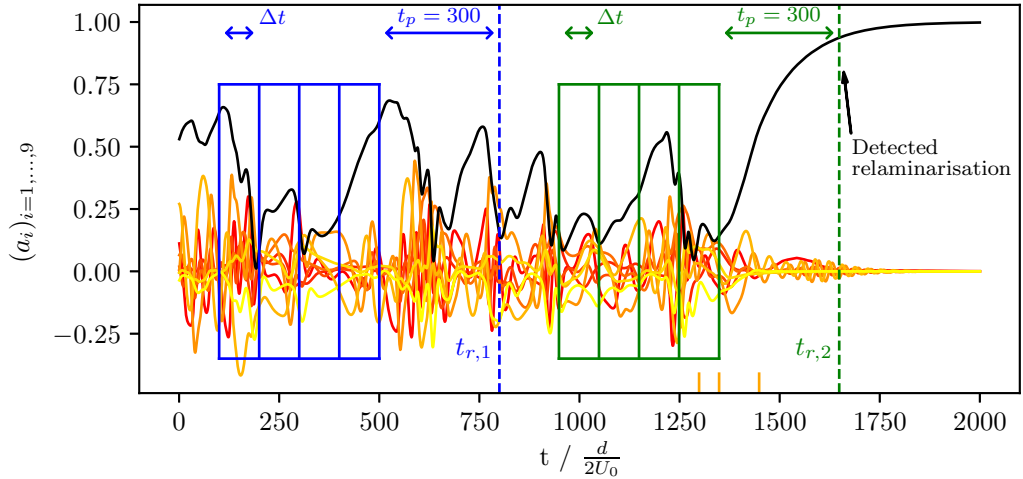}}
	\caption{
		Time series of the nine spectral coefficients $a_i$ in eq.~\eqref{eq:nmm_expansion}, with
		the laminar coefficient $a_1$ shown in black and modes $a_2$ to $a_9$ are shown in red to yellow. The dashed green line represents
		the threshold between turbulent
		and laminar dynamics as defined by an energy threshold
		on the deviations of the laminar profile
		$E_l = 5 \times 10^{-3}$, see eq.~\eqref{eq:turbulent_energy}.
		The number of snapshots per training sample
		is set to $N_s=5$, which are $\Delta t$ apart. 
		The temporal spacing is set to $\Delta
		t=100$ in this example for visual purposes only,
		$\Delta t=3$ is used in all calculations with $N_s > 1$.
		The
		short orange vertical lines mark prediction time horizons of
		$t_p=\{200,300,350\}$ for visual guidance, see Section 4 for further details.
	}
	\label{fig:NMM_dynamics}
\end{figure}

The nine ordinary differential equations that comprise the NMM are solved with
an explicit Runge-Kutta method of order 5 
\citep{dormand1980family}
with a fixed time step, using
Scipy \citep{2020SciPy-NMeth} with Python. The time step for the integrator is
set to $dt=0.25$ for all simulations and we use a simulation domain of size
$[0,4\pi] \times [-1,1] \times [0,2\pi]$ in units of $d/2$.
Since we later train ML models to predict the relaminarisation events, a
Reynolds number of $\R=250$ is chosen in order to reduce waiting times for
relaminarisation events, as the mean turbulent lifetime increases very rapidly
with Reynolds number.
Figure~\ref{fig:NMM_dynamics} presents a time series of $a_1(t), \hdots,
a_9(t)$ representative of a relaminarisation event in the NMM. After irregular
fluctuations, eventually the coefficients $a_2(t), \hdots, a_9(t)$, pertaining to all but the laminar mode, decay.  In contrast, the
coefficient $a_1(t)$ of the laminar mode, shown in red, asymptotes to unity.
The chaotic regions of the dynamics of the NMM are characterised by a Lyapunov
time of $t_{L} \approx 60$. 
The Lyapunov time is the inverse of the largest Lyapunov exponent \citep{ott2002chaos} and corresponds to 
the time after which initially infinitesimally close phase-space trajectories become separated by an $L_2$-distance of $e$, Euler's number.

\section{\label{sec:ML_and_SHAP}Machine learning and SHAP values}

\subsection{\label{sec:ML}XGBoost}

A gradient boosted tree model is used as ML model for making the relaminarisation predictions. Specifically, the \emphasize{XGBoost} \citep{chen2016xgboost} implementation of a boosted tree model in Python is utilised to benefit from its fast implementation for very large datasets. XGBoost is known for its high performances on ML tasks such as high energy physics event classification, massive online course dropout rate predictions and other dedicated real-life ML competition tasks \citep{chen2016xgboost}. Additionally, XGBoost-based classifiers benefit from fast implementations of SHAP value computations \citep{lundberg2018consistent} that will be used in Sec.~\ref{sec:results.explanation} to explain the trained ML model.

Boosting methods belong to the class of ensemble methods \citep{hastie2009elements}. These methods use an ensemble of weak learners, i.e. models that by themselves are not very powerful, to make predictions. The mathematical details of boosted trees and XGBoost can be found in Appendix~\ref{sec:appendix.boosted_trees}.

\subsection{\label{sec:SHAP}SHAP values}

While ML models might show good prediction performances given a task, it is not
per se clear which relations have been learned and led to this good
performance.  Complex and well performing ML models come at the cost of being
difficult to be interpreted and inspected. Hence, traditionally less performing
methods, such as linear models, were deployed for the sake of being easier to
be interpreted. Recent advances in \emphasize{explainable AI} attempt to work
on the understanding of well-performing and complex ML models - including model
agnostic explanation techniques and model-specific explanation techniques - to
benefit from high prediction performances as well as explainable models.

One recent method that enables complex models to be interpreted are
\emph{SHapley Additive exPlanations} (SHAP) values. SHAP values unify recently
developed explainable AI methods such as the LIME \citep{ribeiro2016lime},
DeepLIFT \citep{shrikumar2017deepLift} and layer-wise relevance propagation
\citep{bach2015lrp} algorithms while also demonstrating theoretically that SHAP
values provide multiple desirable properties. Additionally, SHAP values can be
evaluated efficiently when using model-specific implementations such as for
XGBoost.  We briefly introduce SHAP values in the following.

SHAP values belong to the class of \emphasize{additive feature explanation models} that explain the ML model output $g$ at sample $\vec{z}\in\mathbb{R}^M$ in terms of effects assigned to each of the features,
\begin{equation}
	g(\vec{z}) = \Phi_0 + \sum_{m=1}^{M} \Phi_m,
\end{equation}
with $M$ as number of features. \cite{lundberg2017unified} define a specific choice of $\Phi_m$ which they coined as \emphasize{SHAP values}. These are based on the game theoretic Shapley values \citep{shapley1953value} and adhere to three desirable properties that make their explanations locally accurate and consistent. The SHAP value for feature $m$ of sample $\vec{z}$ for model $g$ are computed as
\begin{equation}
	\Phi_m(g, \vec{z}) = \sum_{S \subseteq S_F \setminus \{m\}} \frac{|S|! (M - |S| - 1)!}{M!} (g(S\cup \{m\}) - g(S))
\end{equation}
with $S$ as subset of features that does not contain the feature $m$ to be explained, $S_F$ the set of all $M$ features and $g(S)$ as model output of feature subset $S$. $\Phi_0$ is determined separately as the average model output by $\Phi_0=g(S=\emptyset)$. 

	Intuitively, SHAP values thereby measure the difference between the
	trained model evaluated including a particular target feature and
	evaluated excluding it, averaged over all feature set combinations that
	do not include the target feature. The prefactor is a symmetric
	weighting factor and puts emphasis on model output differences for
	feature subsets $S$ with either a small number of features or a number
	close to $M$. Hence, the model output difference that stems from
	removing the target feature is considered particularly relevant when
	there is either a small or a large number of features in the feature
	set $S$ that is considered.

The model $g$ evaluated on a feature subset $S$, $g(S)$, is technically challenging as a model is trained on a fixed number of features. $g(S)$ is realised by a conditional expectation value that conditions on the feature values of $\vec{z}$ that are present in feature subset $S$,
\begin{equation}
	g(S) = \mathbb{E}[g(\hat{\vec{z}}) | \hat{\vec{z}}=\vec{z}_S].
	\label{eq:shap.conditional_expectation}
\end{equation}
This avoids the technical difficulty of evaluating a readily trained model on a subset of features.

The SHAP value property of local accuracy ensures that the sum of the SHAP values for the explained sample $\vec{z}$ corresponds to the difference between the model output for that sample, $g(\vec{z})$, and the mean prediction of the model, $\langle g(\tilde{\vec{z}}) \rangle_{\tilde{\vec{z}}}$,
\begin{equation}
	\sum_{m=1}^{M} \Phi_m(g, \vec{z}) = g(\vec{z}) - \langle g(\tilde{\vec{z}}) \rangle_{\tilde{\vec{z}}}.
	\label{eq:shap.additivity}
\end{equation}
Hence, the sum over all SHAP values is equal to the difference between model output and mean model prediction.

We use a fast implementation of SHAP values for tree ensemble models by \cite{lundberg2018consistent}. While Eq.~\eqref{eq:shap.conditional_expectation} is typically evaluated by an integration over a background dataset, the fast tree-specific algorithm incorporates the tree structure by omitting all paths that are not compatible with the conditional values $\vec{z}_S$.

	While SHAP values provide per-sample contributions for each feature, a
	typical task is to assign each feature $m=1,\dots,M$ an importance for
	the model predictions. A common approach is to average the absolute
	SHAP values over all samples in the dataset
	\citep{molnar2020interpretable}. The average ensures a statistical
	statement about the SHAP values and removing the sign from the SHAP
	values ensures that positive and negative contributions to the ML model
	output are accounted for equally.

Additionally to the classical SHAP values presented above, there exist SHAP interaction values \citep{lundberg2020local2global} that capture the contributions of feature interactions to the ML model output by generalising the classical SHAP values to combinations of features. Consequently, each sample is assigned a matrix of SHAP interaction values that are computed as
\begin{equation}
	\Phi_{m,n}(g, \vec{z}) = \sum_{S \subseteq S_F \setminus \{m,n\}} \frac{|S|! (M - |S| - 2)!}{2 (M-1)!} \big(g(S\cup \{m,n\}) - g(S\cup \{n\})-[g(S\cup\{m\})-g(S)] \big)
\end{equation}
for $m\neq n$ and
\begin{equation}
	\Phi_{m,m}(g, \vec{z}) = \Phi_m(g,\vec{z}) - \sum_{n\neq m} \Phi_{m,n}(g, \vec{z}).
\end{equation}
Setting $\Phi_{0,0}(g, \vec{z})$ to the average output of $g$, one obtains a similar property as for the classical SHAP values in eq. \eqref{eq:shap.additivity}, namely the additivity property
\begin{equation}
	\sum_{m=0}^{M} \sum_{n=0}^{M} \Phi_{m,n}(g, \vec{z}) = g(\vec{z}).
\end{equation}
Also for these SHAP interaction values we use a fast implementation for tree ensembles \citep{lundberg2020local2global}.

\section{\label{sec:results}Results}

Before studying the inner workings of the ML model, a well-performing model
needs to be trained on relaminarisation events. This section defines the fluid
dynamical classification task and presents the achieved results with a XGBoost
tree followed by their explanation with SHAP values.

The prediction of the relaminarisation events a time $t_p$ ahead is considered
a supervised binary classification problem in ML \citep{bishop2006pattern}.
Supervised tasks require the training data to consist of pairs of input and
target outputs, commonly called $\vec{z}$ and $y$, respectively. Here, the
input data consists of a number $N_s$ of nine dimensional vectors of
spectral coefficients $\vec{a} = (a_1, \hdots, a_9)$ from the flow model introduced in
sec.~\ref{sec:fluid_dynamical_system}. The output is a binary variable encoded
as $1$ and $0$ that contains information on whether the flow corresponding to
the input spectral coefficients relaminarised a time $t_p$ ahead or not,
respectively.

The training data is acquired by forward simulation of the flow model. A
single fluid simulation is initialised with a random nine dimensional initial
condition, with initial amplitudes uniformly distributed according to $U(-0.25,0.25)$,  
and integrated for $4000$ time units. After removing a transient
period of $200$ time units to ensure that the dynamics has reached the
attracting phase space region, training samples for each of the two classes are
extracted from the trajectory. This process of starting forward simulations of
the fluid model and the subsequent extraction of training data is repeated
until enough training samples have been obtained.

The training data comprises of $N_t=10^6$ training samples, half of which belong to the class of samples that relaminarise and that do not relaminarise, respectively. The balanced test dataset is separate from the training dataset and consists of $N_v=10^5$ samples that have not been used for training purposes.

The extraction of training samples from a trajectory is based on the
classification of the trajectory in turbulent and laminar regions. For that,
the energy of the deviation from the laminar flow of each of the velocity fields $\vec{u}(\vec{x}, t)$ in
the trajectory is computed as
\begin{equation}
	E(t) = \langle \vec{u}(\vec{x}, t) - \vec{U}(x_2), \vec{u}(\vec{x}, t) - \vec{U}(x_2) \rangle = \sum_{i=1}^{9} \Big(a_i(t) - \delta_{1, i}\Big)^2,
	\label{eq:turbulent_energy}
\end{equation}
using the spectral expansion coefficients $a_i(t)$ at each time step and the
orthonormality of the basis functions in the last equality.
To classify the
trajectory in turbulent and laminar sections, an energy threshold $E_l =
5\cdot 10^{-3}$ is set. Hence, a velocity field $\vec{u}$ is classified
according to the binary variable
\begin{equation}
	c(\vec{u}) =
	\begin{cases}
		1, & \text{if } E(\vec{u}) \le E_l \\
		0, & \text{else}
	\end{cases}
	\label{eq:energy_criterion}
\end{equation}
with $c = 0$ denoting the class of samples that do not relaminarise $t_p$
time steps ahead and class $1$ denoting those that do relaminarise. The value
for $E_l$ is chosen based on empirical tests that have shown no return to
chaotic dynamics after a trajectory reached a velocity field with energy $E_l$.

Using the classification $c(\vec{u}(t))$ of a trajectory $\vec{u}(t)$, the
training data acquisition is characterised by the prediction horizon $t_p$ and
the number of flow fields $N_s$ that make up one sample. To construct a single
training sample from a trajectory, a random point $t_r$ in the trajectory is
chosen to serve as target point. Its classification label $c(\vec{u}(t_r))$ is
used as training data target output $y$. The input data $\vec{z}$ is obtained
by using $N_s$ equally spaced spectral coefficients preceding the chosen time
point about $t_p$, i.e. at $t_r - t_p$. Hence, a single training sample is
extracted as
\begin{equation}
	(\vec{z}, y) = \Big( \Big[\vec{a}(t_r - t_p - (N_s-1) \Delta t),
	       \dots,
	       \vec{a}(t_r - t_p - 0 \Delta t)\Big],
	       \quad c(\vec{u}(t_r))\Big),
\end{equation}
with the temporal spacing between subsequent snapshots for one training sample
$\Delta t$. We gauged $\Delta t=3$ to the dynamics of the flow model in order
to capture sufficient dynamical detail. Finally, the temporal positions $t_r$
are spread randomly in turbulent regions to obtain samples for class $0$ and
specifically placed at the laminar transition to obtain samples for class $1$.

Figure~\ref{fig:NMM_dynamics} shows the training data acquisition process based
on an example trajectory with one randomly chosen time $t_{r,1}$ to obtain a
sample for class $0$, coloured in blue, and another time $t_{r,2}$ set at the
laminar transition to obtain a sample for class $1$, coloured in green. The
short orange vertical lines mark the prediction time	horizons of
$t_p=\{200,300,350\}$ for visual guidance. The large value of $a_1$ for
$t_p=200$ demonstrates why this prediction horizon serves as validation case.
After training, the ML classifier can be given a set of $N_s$ points equally
spaced with $\Delta t$ and predict whether the flow described by this data 
will relaminarise after a time integration of $t_p$ time units.

\begin{figure}
	\centering
	\includegraphics[scale=1]{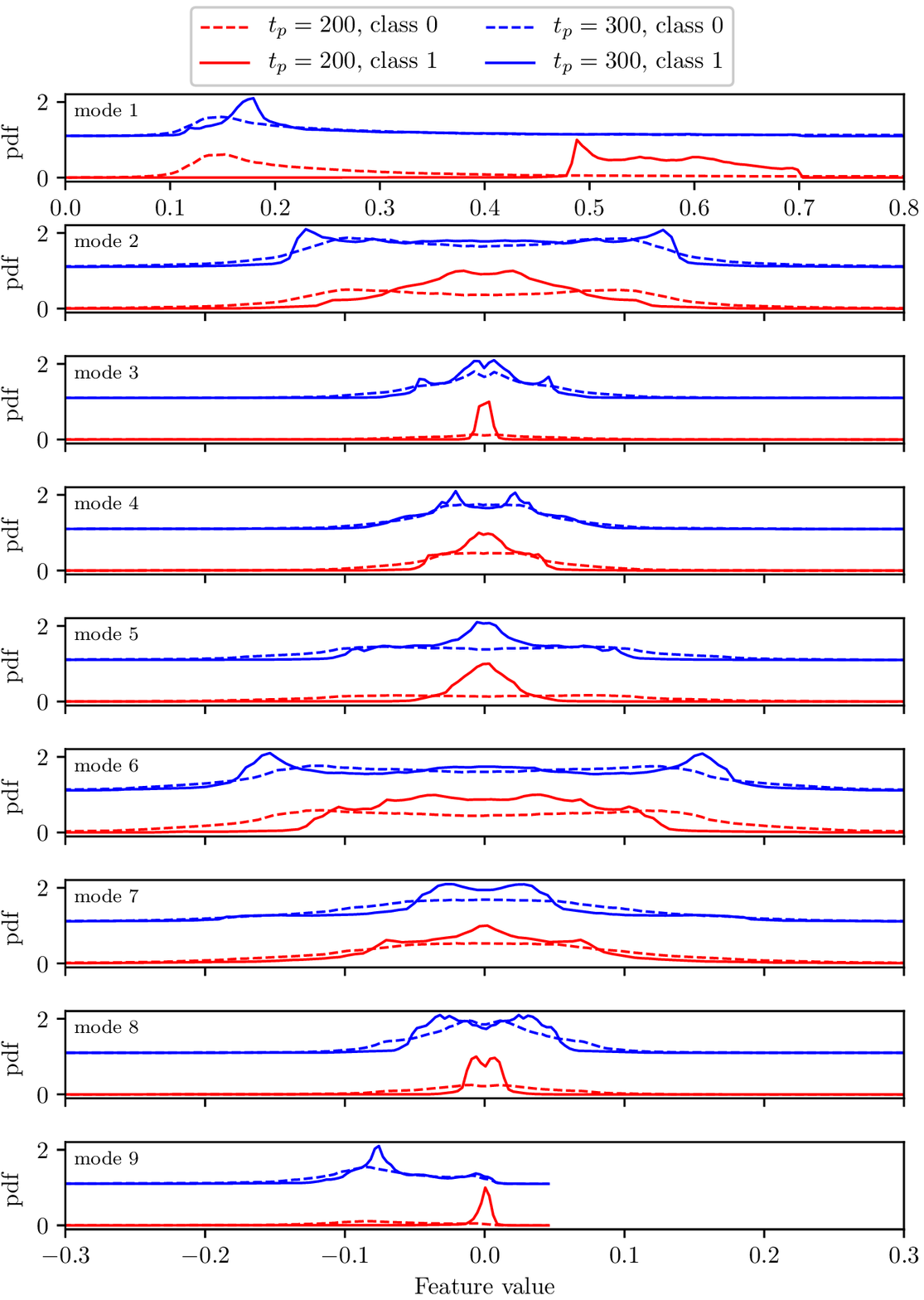}
	\caption{Normalised training data distributions of modes $1$ to $9$.} $t_p=300$ is shifted upwards for visual purposes. 
	Class $1$ ($0$) corresponds to samples that do (not) relaminarise after $t_p$ time steps.
	\label{fig:histogram_training_data_distributions}
\end{figure}

It is good practice to analyse the training data prior to training classifiers
on it. We pick $N_s=1$ and visualise the training data distributions of the
nine spectral expansion coefficients for $t_p\in\{200, 300\}$, see
Fig.~\ref{fig:histogram_training_data_distributions}. The distributions for the two classes $0$ and $1$ become statistically less distinguishable for increasing $t_p$, requiring the ML model to learn per-sample correlations to perform well. It is observed that the classes for $t_p=200$ can
be distinguished from a statistical point of view already. This is because the
prediction horizon is not large enough to move the samples off the slope of the
laminar transition as indicated by the rightmost orange bar in
Fig.~\ref{fig:NMM_dynamics}. The prediction horizon of $t_p=300$, on the other
hand, is large enough to forbid sample classification through simple
statistical properties because the histograms of both classes mostly overlap.
Hence, $t_p=200$ is considered a benchmark case as the prediction performance
is expected to be high because the large laminar mode is sufficient for the
classification.

\subsection{\label{sec:results.classification_task}Prediction of relaminarisation events}

The hyperparameters of the gradient boosted tree are optimised using a
randomised hyperparameter search strategy. The strategy chooses the
hyperparameters from predefined continuous (discrete) uniform distributions
$U_c(a, b)$ ($U_d(a, b)$) between values $a$ and $b$ and samples a fixed number
of draws. We draw $100$ hyperparameter combinations according to the
distributions
\begin{equation}
\begin{split}
	h_{NE}  & \sim U_d(500, 1500), \\
	h_{MD}  & \sim U_d(1, 50), \\
	h_{MCW} & \sim U_d(1, 15), \\
	h_{GA}  & \sim U_c(0, 5), \\
	h_{SS}  & \sim U_c(0.5, 1), \\
	h_{CBT} & \sim U_c(0.5, 1), \\
	h_{LR}  & \sim U_c(0.001, 0.4).
\end{split}
\end{equation}
We verified for $t_p=300$ that $100$ draws cover the hyperparameter phase space
sufficiently well by drawing $200$ hyperparameter combinations to show that
this leads to similar prediction performances as for $100$. The hyperparameters
that are found by the randomised hyperparameter search are listed in
Appendix \ref{sec:appendix.optimal_hyperparameters}.

\begin{figure}
	\centering
	\includegraphics[scale=1]{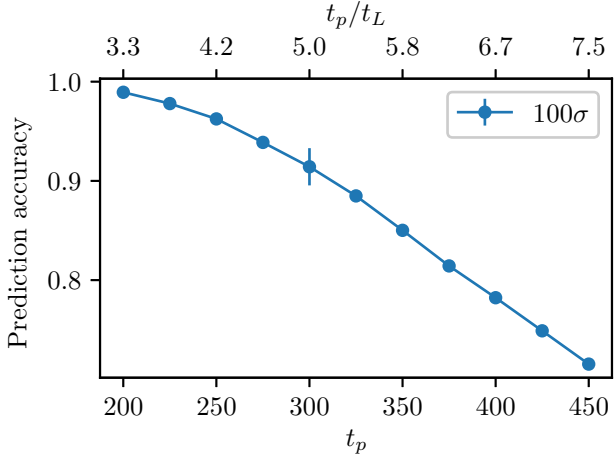}
	\caption{
		Prediction performance of the trained classifier
		against temporal prediction horizon $t_p$. The error bar for
		$t_p=300$ shows $100$ standard deviations $\sigma$ for visualisation purposes and
		demonstrates the robustness of the results. It has been obtained by
		training the classifier on training datasets based on different initial random number generator seeds.
		}
	\label{fig:prediction_performance}
\end{figure}

The prediction performance, measured on a test dataset, for $N_s=1$ decays with increasing $t_p$, as
expected on account of the intrinsic chaotic dynamics of the flow model
\citep{lellep2020using,moehlis2005periodic,moehlis2002models}, see Fig.~\ref{fig:prediction_performance}. Nevertheless,
the prediction performance is around $90\%$ for $5$ Lyapunov times
\citep{bezruchko2010extracting} in the future and is, thereby, sufficiently
good for the subsequent model explanations by SHAP values. Calculations for
different values of $N_s$ verify that the prediction performance only varies
marginally for one exemplary set of hyperparameters. This is to be expected
based on the deterministic nature of the dynamical system and its full
observability. Hence, we here focus on $N_s=1$, which means that the classifier
does not get dynamical information but only a single spectral snapshot of the
flow field. This reduces the computational cost for the subsequent model
explanation by SHAP values.

The prediction horizon $t_p=200$, indeed, corresponds to the benchmark case
where the laminar mode is supposed to be the only relevant indicator for
relaminarisation and $450$ corresponds to the case beyond which the ML model
cannot predict reliably due to the chaotic nature of the system
\citep{lellep2020using,moehlis2005periodic,moehlis2002models}.

\begin{figure}
	\centering
	\includegraphics[width=\columnwidth]{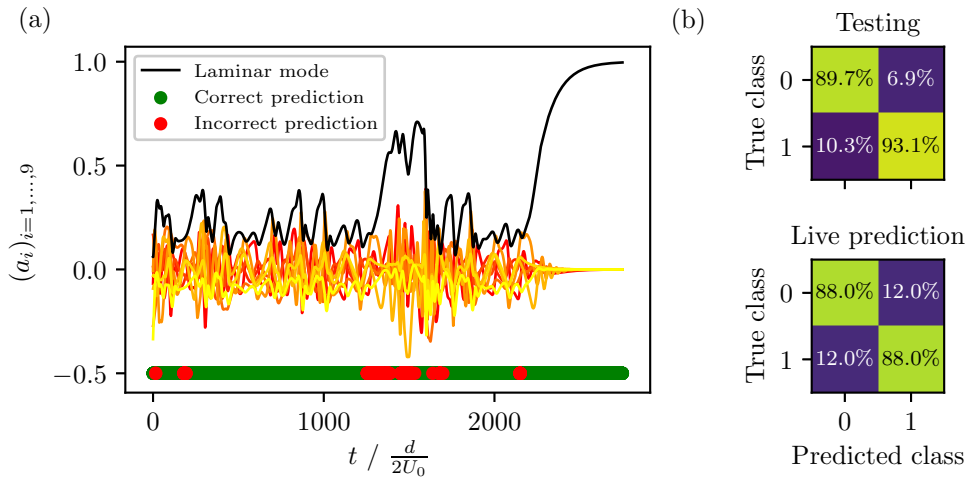}
	\caption{
		Classifier applied in parallel to fluid simulation. (a) Time series with indicated prediction output. The mode corresponding to the laminar profile, $a_1$, 
		is shown in black and modes $a_2$ to $a_9$ are shown in red to yellow.
		(b) Compared normalised confusion matrices of the model evaluated
		on the test dataset (top) and during the live prediction (bottom). The normalisation is required to compare both confusion matrices because of the class imbalance between model testing and live prediction. See main text for more details.
	}
	\label{fig:classification_task.results.live_predict}
\end{figure}

Lastly, to demonstrate the performance of the machine learning model also for
applied tasks, the model is applied in parallel to a running fluid
simulation. Figure ~\ref{fig:classification_task.results.live_predict}(a) shows the on-line prediction of one simulated trajectory. The horizontal bottom bar indicates whether the prediction of the 		classifier has been correct (green) or incorrect (red). We collected statistics
over $1000$ trajectories to quantify how well the model performs on an applied
task instead of the test dataset. As shown in
Fig.~\ref{fig:classification_task.results.live_predict}(b), the model
performance for the on-line live prediction is with around $90\%$ true positives and true negatives as well as around $10\%$ false positives and false negatives comparable to the performance on
the test dataset in terms of the normalised confusion matrices of the predictions. The normalisation of the confusion matrices is necessary to account for the substantial class imbalance in the data pertaining to the live prediction and to, thereby, make the performances on the two tasks comparable.

The results demonstrate that the ML model performs sufficiently well despite
the intrinsic difficulty of predicting chaotic dynamics. Next, we turn
towards the main contribution of this work. There, we use the trained XGBoost
ML model as high-performing state-of-the-art ML model together with SHAP values
to identify the most relevant physical processes for the relaminarisation
prediction in shear flows in a purely data-driven manner.

\subsection{\label{sec:results.explanation}Explanation of relaminarisation predictions}

Since SHAP values offer explanations per sample and there are many samples to
explain using the test dataset, two approaches may be taken: First,
a statistical statement can be obtained by evaluating the histograms of SHAP
values of all explained samples. Second, live explanations of single samples can
be calculated, similar to what we demonstrated previously in
Sec.~\ref{sec:results.classification_task} with live predictions of
relaminarisation events. This work focuses on the former of the two
perspectives and notes the potential of the latter approach for future work in
Sec.~\ref{sec:conclusions}.

\begin{figure}
	\centering
	\includegraphics[scale=1]{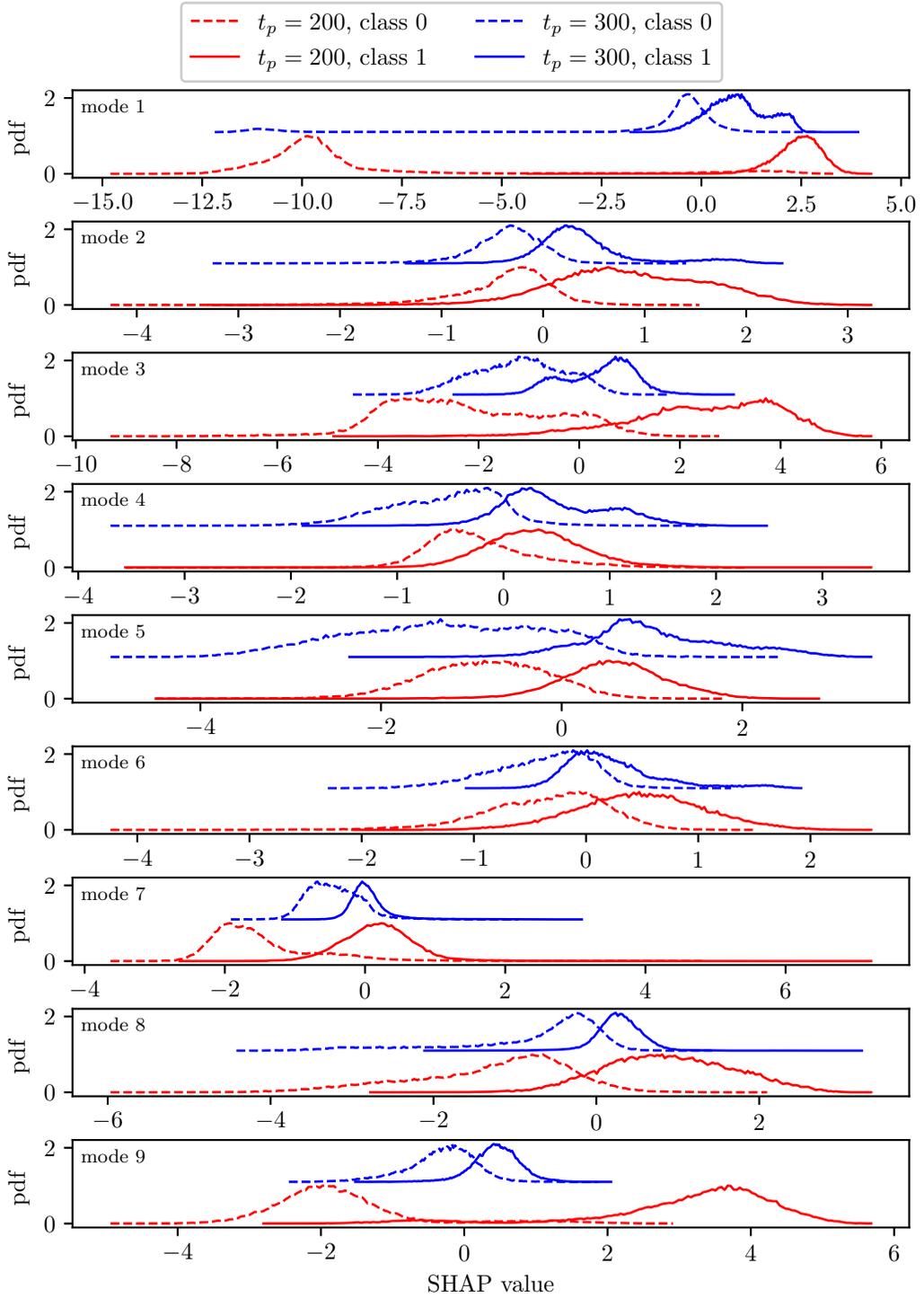}
	\caption{Normalised SHAP value distributions of modes $1$ to $9$ for the $10^5$ test samples, $t_p=300$ is shifted upwards for visual purposes. Class $1$ ($0$) corresponds to samples that do (not) relaminarise after $t_p$ time steps.}
	\label{fig:histogram_shap_values}
\end{figure}

The statistical evaluation shows bi-modal SHAP value distributions, see
Fig.~\ref{fig:histogram_shap_values}. Each class corresponds to one of the modes,
emphasising that the model learned to distinguish between the two classes
internally as the two classes are explained differently.

\begin{figure}
	\centering
	\includegraphics[scale=1]{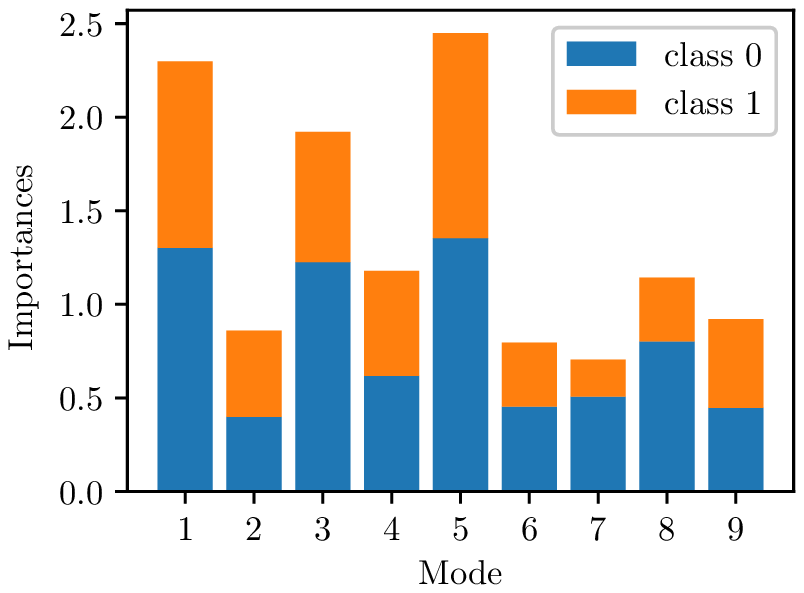}
	\caption{
		The mean absolute SHAP values as
		distinguished by the underlying class for $t_p=300$. Class $1$ ($0$) corresponds to samples that do (not) relaminarise after $t_p$ time steps.
		}
	\label{fig:mean_SHAP_values}
\end{figure}

\begin{figure}
	\centering
	\includegraphics[width=\columnwidth]{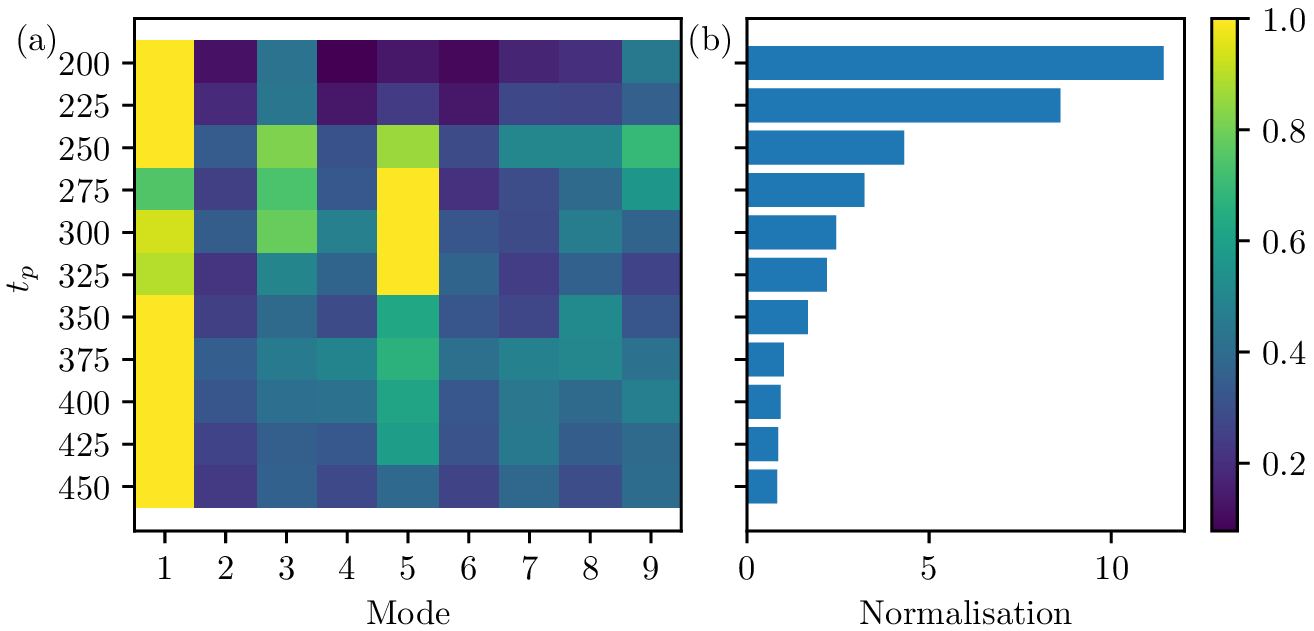}
	\caption{
		Feature importances as measured by mean absolute SHAP values.
		(a) The feature importances normalised separately
		for each $t_p$ along its row to show the hierarchy of mode importance. (b)
		Normalisation constants used in (a). To convert the normalised values shown in panel (a) 
		to their absolute counterparts, each row would need to be multiplied by the corresponding normalisation shown in panel (b).
	}
	\label{fig:explanations.results.feature_importances}
\end{figure}

From Eq.~\eqref{eq:shap.additivity} follows that the model output $g(\vec{z})$
is made up of the SHAP values $\Phi_m(g, \vec{z})$. The multi-modality of the
SHAP values conditional on the class means therefore that the feature
contributions to the final output differ for both classes.
Figure~\ref{fig:mean_SHAP_values} shows the average absolute SHAP values per
class over all explained samples for $t_p=300$ and thereby quantifies the differences in mode
importance for the prediction of the two classes
\citep{molnar2020interpretable}. Hence, the figure demonstrates that modes $1$, $3$ and $5$ are
the three most important modes.
Feature importances are evaluated by computing the absolute SHAP value mean
for different prediction times. This is shown in
Fig.~\ref{fig:explanations.results.feature_importances} for a range of $t_p$.

The robustness of these results has been validated by two means: First, the
SHAP values of $t_p=300$, which is neither a trivial task nor suffers from bad
prediction performance at a prediction performance of $91\%$, are recomputed
for a second set of randomly acquired training data by using a different
initial training data seed. Not only do the minute fluctuations in
Fig.~\ref{fig:prediction_performance} indicate the close similarities between the
results but also the SHAP value histograms are similar (data not shown).
Second, the XGBoost model with the optimal hyperparameters is retrained on a
subset of features that are chosen according to the feature importances derived from the SHAP values. The computations
confirm that the basis functions - which here have a clear correspondence to
physical features and dynamical mechanisms -
identified as most important features by the
SHAP values lead to the largest training performance of all subsets tested.
Also, the least important modes lead to the lowest training performance.
Lastly, the baseline of all modes consistently achieves the largest prediction
performance. Additional to the few tested feature subset combinations, all
$\binom{9}{3}=84$ combinations to pick 3 out of the 9 features have been
evaluated for $t_p=300$. For these subsets, the prediction accuracy varies
between $65\%$ and $80\%$,  with the combination of the features with the
largest SHAP values, $(1,3,5)$, leading to the maximal prediction accuracy (not shown).

\begin{figure}
	\centering
	\includegraphics[scale=1]{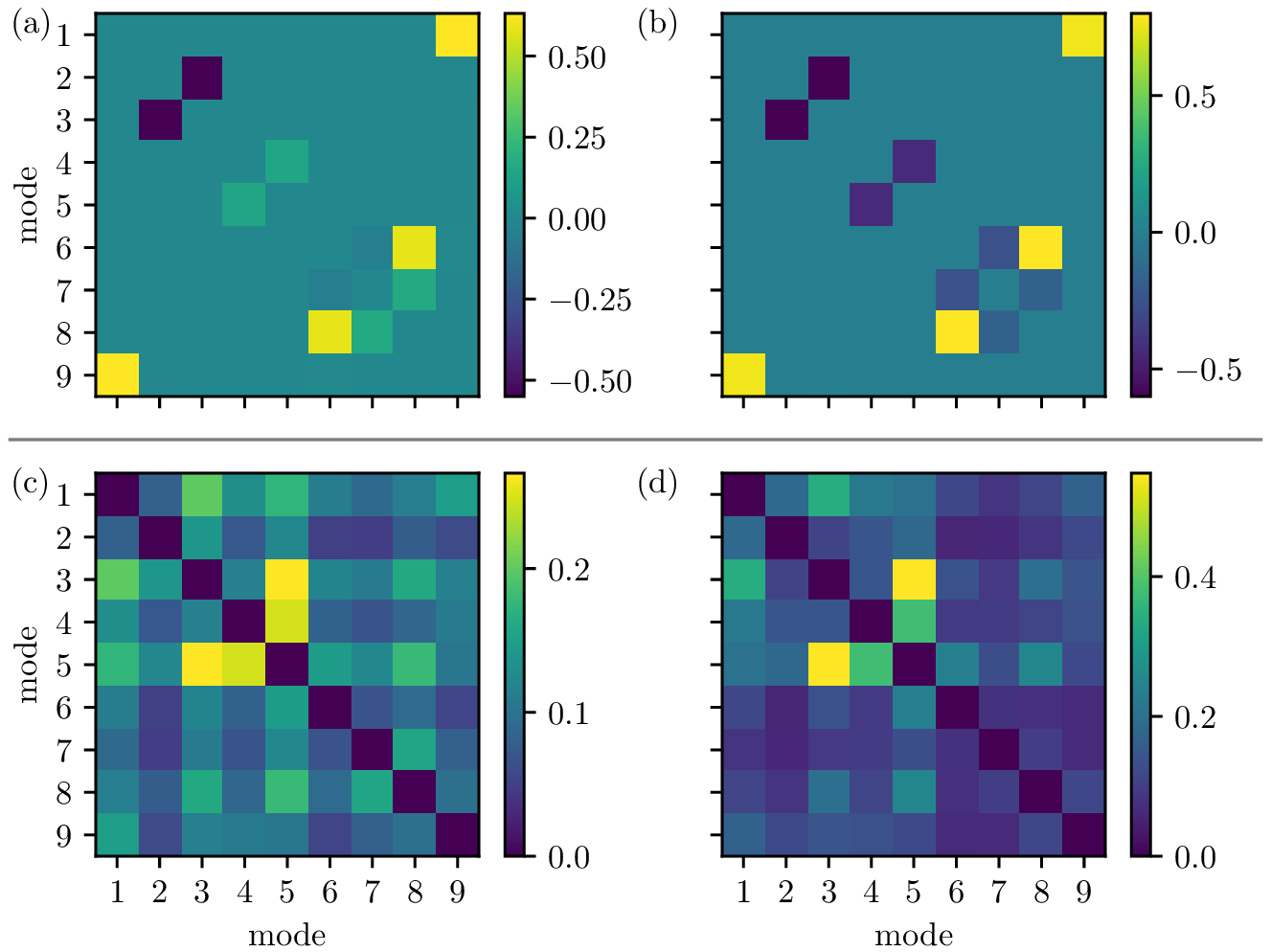}
	\caption{
		(a) and (b): Correlation matrices of training data for classes
		0 and 1, respectively. Class $1$ ($0$) corresponds to samples that do (not) relaminarise after $t_p$ time steps. (c) and (d): Mean absolute SHAP
		interaction values of the first $N_v/10=10000$ validation
		samples for classes 0 and 1, respectively. As self-correlations encoded in the diagonal elements do not convey useful information, the diagonal elements have been set to zero for all subfigures for presentational purposes.
	}
	\label{fig:results.interaction_matrices}
\end{figure}

To appreciate the concept of SHAP values, it is instructive to consider
correlation matrices of the training data as shown in
Fig.~\ref{fig:results.interaction_matrices}(a,b) for classes 0 and 1, respectively.
A few observations can be made from the data. First, the correlation matrices
belonging to two classes are remarkably similar, demonstrating that correlations alone are
not sufficient to distinguish between the classes. Here, we note that correlations only
capture linear relations between random variables. 
The only difference is that modes 4 and 5 positively correlate in class 0, while they
correlate negatively in class 1, and similarly for modes 7 and 8.
When comparing the correlation matrices with the mode coupling table or
with the amplitude equations in \citet{Moehlis2004} we observe that strongly
correlating modes couple via either the laminar profile (mode 1) or its
deviation in streamwise direction (mode 9).  The strong negative correlations
between modes 2 and 3, and strong positive correlations between modes 6 and 7,
which occur for both classes, can be made plausible by inspection of the
evolution equation of the laminar profile. The nonlinear coupling only extract
energy from the laminar flow if the amplitudes of modes 2 and 3 have the
opposite sign, and those of modes 6 and 8 are of the same sign as products of
these mode pairs occur in the evolution equation of the laminar profile. In
other words, in order to obtain unsteady dynamics modes 2 and 3 must be mostly
of opposite sign while 6 and 8 must mostly have the same sign. In this context
we note that the amplitude of the laminar mode is always positive, as can be
seen from the top panel of
Fig.~\ref{fig:histogram_training_data_distributions}.

Secondly, we consistently find correlations between modes identified as significant for the
prediction and irrelevant modes. For instance modes one and nine correlate, and so do mode two and three,
four and five. Thirdly, modes that are considered significant do not correlate.
The latter two points highlight the game-theoretic structure of SHAP.
For example, as the fluctuations of the coefficients pertaining to modes two and three are
strongly correlated, it is sufficient for the classifier to know about one of them.
We will return to this point in Sec.~{\ref{sec:interpretation}.}

To elucidate second-order interaction effects further, the SHAP interaction
values \citep{lundberg2020local2global} are computed, see
Fig.~\ref{fig:results.interaction_matrices}(c,d).  The overall bar heights
denote the mode importance across both classes, the coloured bars distinguish
between both classes. Interactions between modes $3$ and $5$ are found to be
strongest for samples from both prediction classes.  In particular modes $7$
and $8$ differ in their importances for the two classes: both are more
important in cases where relaminarisation does not occur.  Interaction effects
between modes $4$ and $5$ are present for both classes, but more pronounced for
samples of class $0$. Generally, interaction effects of samples of class $1$
are stronger than for those of class $0$.

The feature importances presented in
Fig.~\ref{fig:explanations.results.feature_importances} show that the laminar
mode is consistently identified as a relevant feature. The shortest prediction
time $t_p=200$ not only comes with a prediction accuracy of $\approx 98\%$, but
the feature importance of the laminar mode is also significantly stronger than
for the other tested prediction horizons. This indicates that this prediction
case can, indeed, be considered a validation case. Within that scope, the
validation succeeded as the statistically significant laminar mode is detected
as most relevant mode.

Increasing the prediction horizons leads to a decrease in the importance metric
for all features, as can be inferred from the observed decrease in
normalisation factors shown in
Fig.~\ref{fig:explanations.results.feature_importances} (b). 
The normalisation constants shown in Fig.~\ref{fig:explanations.results.feature_importances}(b) are computed as
$N_i=\sum_{j=1}^{9} \Phi_{i,j}$, where $\Phi_{i,j}\in\mathbb{R}$ denotes the SHAP 
value of mode $j \in \{1, \hdots, 9\}$ for a prediction
time $t_p^{(i)}$, the superscript $i$ enumerating the sampled prediction times $t_p \in \{200, 225, \dots, 450\}$.
Figure~\ref{fig:explanations.results.feature_importances}(a) thus presents $\Phi_{i,j}/N_i$.
The observed decrease in normalisation factors with increasing $t_p$ 
indicates,
together with declining prediction performance, that sufficiently
well-performing classifiers are required to enable the subsequent explanation
step.

\subsection{Interpretation}
\label{sec:interpretation}
Throughout the prediction horizons, $\bm{u}_1$, $\bm{u}_3$ and $\bm{u}_5$ are consistently
considered important. These modes represent the
laminar profile, the streamwise vortex and a spanwise sinusoidal linear instability of the streak
mode $\bm{u}_2$, respectively. Streamwise vortices and streaks are a characteristic feature of wall-bounded shear flows \citep{holmes2012turbulence,schmid2002stability,bottin1998discontinuous,Hamilton1995}.
Alongside the laminar profile and its linear instabilities $\bm{u}_4-\bm{u}_7$,
they play a central role in the self-sustaining process, the basic mechanism
sustaining turbulence in wall-bounded shear flows. The importance of the streamwise vortex $\bm{u}_3$
increases with prediction horizon and decreases from $t_p\approx 300$ onwards, where the
prediction accuracy begins to fall below $90\%$.

The streak mode itself appears to be irrelevant for any of the predictions,
which is remarkable as it is, like the streamwise vortex, a representative feature of
near-wall turbulence. Similarly, its instabilities, except mode $\bm{u}_5$, are not of importance
for the classification.
For the shortest prediction time, that is, for the validation case $t_p =
200$, mode $\bm{u}_5$ does not play a decisive role either,
which is plausibly related to the SSP being significantly weakened close to a
relaminarisation event. This rationale, of course, only applies to data samples
in class 1, where relaminarisation occurs.  Like the vortex mode $\bm{u}_3$, the spanwise
instability mode $\bm{u}_5$ increases in importance
with prediction horizon except for the longest prediction horizon,
which again can be plausibly explained by a more vigorous SSP further away from a relaminarisation event.
Since $\bm{u}_1$ and $\bm{u}_3$ are translation-invariant in the streamwise direction,
a mode with $x$-dependence should always be recognised, as the SSP
cannot be maintained in two dimensions. The dominance of $\bm{u}_5$ over any of the
other instabilities may be related to its geometry resulting in a stronger shearing and thus a
faster instability of the streak.

Apart from modes directly connected with the SSP, the deviation of the mean profile from
the laminar flow, $\bm{u}_9$, is also recognised as important for $t_p = 200$ and
$t_p = 250$. Turbulent velocity field fluctuations are known to alter the mean
profile. In extended domains, where turbulence close to its onset occurs in a
localised manner, localisation occurs through turbulence interacting with and
changing the mean profile. The mean profile of a puff in pipe flow, for
instance, is flatter than the Hagen-Poiseuille profile towards the middle of
the domain, which decreases turbulence production \citep{vanDoorne2009,hof2010twopuffs,barkley2016theoretical}.

Now the question arises as to if and how the
information SHAP provides concerning the mode importance ranking 
can be connected to the equations of motion and what can be learned from this. In particular, 
concerning strongly correlated modes, it is instructive to understand why a particular mode 
is favoured.
For modes two and three, the mode coupling table of \citet{Moehlis2004} again gives
some indications.  Mode three (the streamwise vortex) generates mode two (the streak)
by advection of mode one (the laminar profile) -- this is the first step of the SSP 
-- or mode 9 (the deviation of the laminar profile).
However, the coupling table is not symmetric, that is, $\bm{u}_2$ cannot generate
$\bm{u}_3$, and $\bm{u}_3$ can only be re-generated through nonlinear interactions
involving either $\bm{u}_5$ and $\bm{u}_6$ or modes $\bm{u}_4$ and $\bm{u}_7$ or $\bm{u}_8$ 
-- this is known as the third and last step in the SSP, where instabilities couple nonlinearly to re-generate
the streamwise vortex. Hence, out of the strongly correlated mode pair $\bm{u}_2$ and $\bm{u}_3$,
the latter should be {\em physically} more significant in the SSP than the former, in the sense that
$\bm{u}_2$ will become active if $\bm{u}_1$ and $\bm{u}_3$ are, but not vice versa. SHAP indeed
identifies $\bm{u}_3$ as significant while $\bm{u}_2$ plays no decisive role in the
prediction. 
A similar conclusion has recently been obtain for the transition to turbulence in flow through a vertically 
heated pipe \citep{marensi2021heated}, where relaminarisation due to buoyancy forces has been connected with a 
suppression of streamwise vortices rather than streaks.

For modes 4 and 5 the situation is more subtle, as both modes can be converted
into each other through advection by the laminar profile. Again considering the mode
coupling table (or the amplitude equations), two points distinguish mode 5 from
mode 4: (a) mode 5 is odd in $x_2$ while mode 4 is even in $x_2$, (b) in interactions with mode 2, 
mode 5
couples to the only fully 3d mode, mode 8 (which is also odd in $x_2$), while
mode 4 does not. A fully fledged SSP should involve 3d dynamics, and the data
distribution of mode 8 shows this clearly for the validation case ($t_p = 200$)
as mode 8 is significantly weakened in class 1 compared to class 0.
Considering the training data distributions of modes 4, 5 and 8, we observe
that that the pdfs of mode 5 differ considerably between class 0 and class 1,
and again mode 5 is suppressed in class 1. In contrast, mode 4 is active in
both classes. Mode 5 thus provides a more direct route to three-dimensional dynamics
from streak instabilites than mode 4 does. 

In summary, the picture that emerges is as follows. For a sustained SSP, the streamwise vortex
must be remain active as only it can generate the streak.  Further to this,
supplying spanwise flow perturbations of odd parity in wall-normal direction should help to
prevent relaminarisation events, while spanwise flow fluctuations connected
with streak instabilities of even parity in wall-normal direction play a minor role in
sustaining the SSP.

\section{\label{sec:SHAP_values_of_DNS}Example - SHAP on data obtained by Direct Numerical Simulation}

In order to demonstrate that the method can be leveraged to larger
fluid dynamics datasets, we now discuss an example where SHAP values
corresponding to the prediction of relaminarisation events are calculated on a
dataset obtained by DNS of minimal plane Couette flow at transitional Reynolds number.  
For this particular example, features are not tied to physical
processes or any modal representation of the flow. Instead, the analysis is carried out on flow-field 
samples, and SHAP is used in conjuction with a now neural-network based classifier 
to provide (a) an indication as to which flow features need to be observed to allow 
an accurate prediction of relaminarisation events, and (b) an interpretation thereof in terms 
of the SSP. 
Specifically to address the latter, and to connect to the results obtained for the SSP within the NMM, 
we ask the classifier to predict relaminarisation events based on the structure of the streaks characteristic for the SSP, 
and we use SHAP to demonstrate that the classifier bases its decisions on data features indicating the presence of 
streak instabilities. To do so, we focus on the streamwise component of the velocity field evaluated at a particular point 
in streamwise direction, that is, the classifier works with 2D data slices showing cross-sections of high- and low-speed streaks.

\subsection{Numerical Experiments}

In order to keep the computational effort to a manageable level commensurate with an example calculation, 
we consider simulations of plane Couette flow at a Reynolds number of 400 in the minimal flow unit, a domain of size $L_1\times
L_2 \times L_3 =1.755\pi \times 2 \times 1.2\pi$ with periodic boundary conditions in stream- and spanwise directions and no-slip boundary conditions in the wall-normal direction, 
the smallest domain that sustains turbulence \citep{Jimenez1991,Hamilton1995,Kawahara2001}.
The calculations have been carried out with a pseudospectral solver provided by {\tt channelflow2.0}  \citep{channelflowpaper2014, channelflowpaper2022} 
using  $n_1 \times n_2 \times n_3 = 16 \times 33 \times 16$ points in streamwise, wall-normal and spanwise directions, respectively, with full dealiasing in stream- and spanwise directions, 
a resolution similar to other, including recent, studies of minimal plane Couette flow \citep{Kawahara2001,vanVeen2011,Lustro2019}.
We generate velocity field data for $5000$ trajectories running for $5000$ advective time units 
and take data samples at an interval of 10 advective time units. The simulations are initialised with randomly perturbed velocity-field samples 
taken at intervals of one advective time unit from a turbulent master trajectory 
\review{and the training data acquisition process is started after a transient of 100 advective time units}.
The criterion for the observation of a relaminarisation event is a cross-flow energy threshold of $10^{-5}$, 
and we verify that the turbulent lifetime distribution is still exponential for this grid size (not shown). 
%

As indicated earlier, a convolutional neural network is trained on the 
streamwise slice at $x_1=0$ of the streamwise $u_1$ component with around $5000$ samples, yielding a spatial sample size of
$10 \times 33$, taking into account truncation to remove aliasing effects. Two 2D convolutional layers with subsequent 2D
max pooling layers are followed by a flattening layer with a dropout layer and
a fully connected softmax layer with two neurons, one for each output class
\citep{Chollet2021}, to establish the NN graph. The size of the snapshots is well within the capabilities of 
NNs, that is, the method can certainly be applied to higher-resolved data.  
The main reason for the choice of resolution here is that the exponential lifetime distribution 
results in having to discard a significant number of trajectories, essentially all those 
where relaminarisation occurred very quickly, in order to ensure that 
the transient from the initial data has passed. 
After training the NN, the SHAP values for samples from the test
dataset are calculated.  We can focus on the SHAP values for class $1$ only, as the
SHAP values for the two classes differ only by a minus sign. 

\subsection{Results}

\begin{figure}
	\centering
	\includegraphics[scale=1]{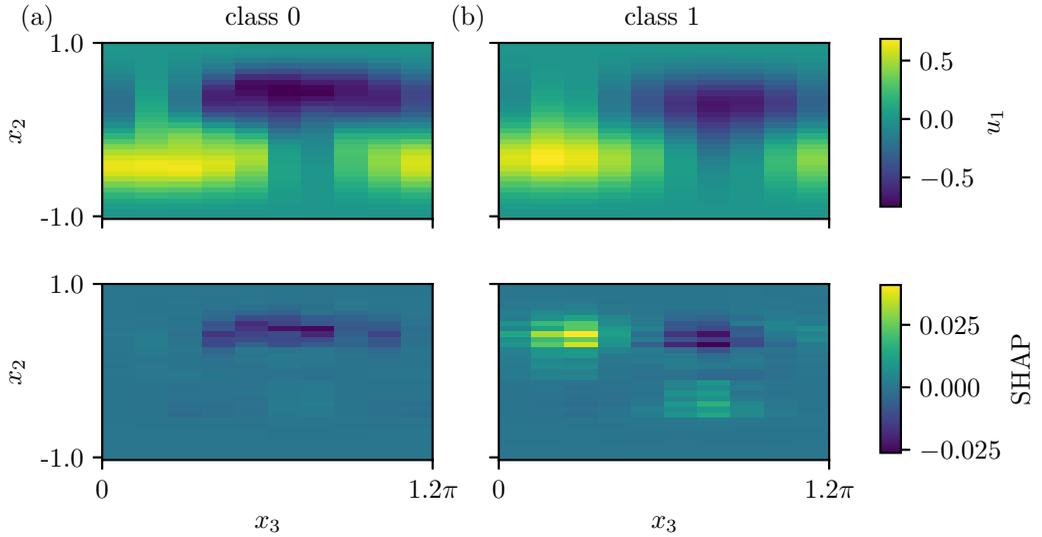}
	\caption{
		Two representative velocity field samples (top) and
		corresponding SHAP values (bottom) for (a) class $0$ and (b)
		class $1$. For the velocity field, \review{streak cross sections, that is} the deviation of the
		streamwise velocity component from the laminar profile
		evaluated at $x_1 = 0$ is shown. 
		\review{As can be seen by comparsion of the top and bottom panels, SHAP uses streak tails, that is 
		regions where the streaks are spatially decaying, for classification towards class $1$ and streak cores, 
		where the velocity is nearly uniform, for classification towards class $0$.} 
	}
	\label{fig:dns_shap_values}
\end{figure}

First, the prediction time $t_p$ is varied between $10$ and $200$ advective time units to obtain the
performance of the convolutional network for tasks of different difficulties
\citep{lellep2020using}. The performance decreases from around $99\%$ prediction
accuracy for $t_p<60$ to around $60\%$ at $t_p=200$ with a performance $>90\%$
for $t_p \leq 130$ (not shown). In what follows, we discuss results obtained for $t_p = 90$, however, results are 
consistent with larger prediction horizons with a prediction accuracy of $>90\%$. 
In Fig.~\ref{fig:dns_shap_values} we present one representative sample for each of the two
classes together with the spatial distribution of SHAP values to illustrate general observations 
that can be obtained from the data.

The streamwise component $u_1$ of velocity-field samples evaluated at $x_1 = 0$ 
always consists of localised regions in the velocity field of alternating small and large magnitudes, 
corresponding to cross sections of low- and high-speed streaks. Samples corresponding to class $0$ feature less uniform 
\review{streak cross sections} than those of class $1$. \review{More precisely, the spatial decay of a streak 
in wall-normal and spanwise directions is less regular for samples of class $0$ than for those of class $1$.} 
This can be seen by comparison 
of the representative visualisations shown in the top panels of Figs.~\ref{fig:dns_shap_values}(a) for class $0$ and
Figs.~\ref{fig:dns_shap_values}(b) for class $1$. 
\review{In what follows, we refer to regions of spatial decay as streak tails.}  

For samples of class $0$, i.e. those that do not relaminarise, the SHAP
values are mostly negative while the SHAP values for samples of class $1$ are
mostly positive. 
Furthermore, for samples of class $0$ SHAP values detect the streak \review{cores, where $u_1$ varies little},
more so for the low-speed rather than the high-speed streak. For class $1$, however, the SHAP values
point towards the tails of the corresponding more uniform streak cross sections. 
The tails of the high-speed streaks are hereby more
pronounced. Interestingly, the tails of the less regular class-$0$ streak cross sections 
slightly contribute towards a classification of those samples to class
$1$ and the inner region of the class-$1$ small velocity regions contribute to
the classification of those samples to class $0$. That is because the tails and
the core look similar to those of the other class, respectively. We can
therefore conclude that the NN uses the streak cores for the
classification towards class $0$ and the streak tails, 
where velocity-field gradients are large, for the classification towards class $1$.

\subsection{Discussion}

The 2D slices used in the training result in the classifier having to predict relaminarisation
events based on the structure of the streaks, but without directly seeing the streak modulations characteristic of 
linear streak instabilities, as these would only be visible in either the full volume or in a wall-normal cross section of the flow. 
Comparing wall-normal cross-sections obtained from samples of classes 0 and 1 {\em a-posteriori}, we find consistently that streak modulations 
are much weaker or absent in class 1 compared with class 0. In conjunction with the results obtained by SHAP, we conclude that 
the classifier bases its decision on proxies for streak modulations corresponding to linear streak instabilities within the SSP.   

\review{
For a relaminarisation event to occur, the SSP must break at some point in its evolution. The SSP consists of three consecutive stages,  
(i) the laminar profile is advected by the streamwise vortex creating streaks,
(ii) the streaks become linearly unstable,
(iii) the linear instabilities couple nonlinearly and re-generate the streamwise vortex. 
Relaminarisation could in principle be related with any of these stages, for instance
with a weakening of streaks or of streamwise vortices or a suppression of streak instabilities. 
Strong streaks are present in both class 0 and class 1 DNS data samples, as can be seen from the visualisations 
of representative data samples shown in Fig.~\ref{fig:dns_shap_values}. 
This is commensurate with the results obtained for the NMM, where 
we found that the streak mode itself is not relevant for the prediction of relaminarisation events. 
That is, a scenario whereby relaminarisation is connected with a suppression of streaks is unlikely. 
A similar observation has been made for buoyancy-induced relaminarisation in a vertically heated pipe \citep{marensi2021heated}. 
In contrast to this, and as discussed above, accurate predictions of relaminarisation events can be 
made based on the presence or absence for proxies for linear streak instabilities.
If they are present, the streaks have a less regular profile and the flow
remains turbulent, if not, it relaminarises. 
We note that the classification task did not involve observables connected with streamwise vortices. 
However, as streamwise vortices are a consequence of streak instabilities, not precursors, a suppression of streak 
instabilities would necessarily result in a suppression of streamwise vortices.  
In summary, the SHAP analysis suggests that the
break-up point in the self-sustaining cycle is connected with a suppression of streak instabilities.
	}

Using 2D slices of a low-resolution DNS near the laminar-turbulent transition
as training data for relaminarisation predictions shows that SHAP values yield
insights into what is important for the prediction in the demonstration carried out here. Specifically, localised SHAP values
can be linked to localised coherent regions of the flow, thereby providing a way to find relevant flow
features for what the convolutional NN classifier uses for the
prediction. Interestingly, regions where velocity-field gradients are large, that is, the tail ends of
the streak cross sections, play a major role in the prediction of relaminarisation events.  

The ability of SHAP to identify localised regions in the flow and isolate important sub-regions therein suggests that 
SHAP values can also identify physical processes that are used by classifiers provided 
a data representation capable of encoding physical, i.e. dynamical, processes is used. 
This would involve passing temporally resolved data, for instance by providing single training samples 
consisting of time-ordered data samples in sub-intervals of a given time series. Additionally, the correlation between input data and
relevance for classifiers as determined by SHAP values could be used by
experimentalists to, for instance, increase measurement resolution where necessary or optimise the location of probes. 

\section{\label{sec:conclusions}Conclusions}

The purpose of this article is to introduce SHAP as an explainable AI method
capable of identifying flow features relevant for the prediction of
relaminarisation events in wall-bounded parallel shear flows.  As a first step
and in order to facilitate a physical interpretation of the SHAP output, we
used a dataset consisting of snapshots generated through forward integrations
of the nine-mode model of \citet{Moehlis2004}, as it is based on the
self-sustaining process and each feature, here basis function, has a clear
physical meaning.
Subsequently, the same classification task is carried out on data obtained
from DNSs of minimal plane Couette flow, where we specifically focus on 
the prediction of relaminarisation event based on the structure of high- and low-speed streaks.
\review{
The feature ranking furnished by SHAP was
interpreted in the context of the SSP, resulting in a clear
distinction between those near-wall features phenomenologically representative of the
flow and those significant for the dynamics of a wall-bounded turbulent flow close to
the onset of turbulence. More specifically, we demonstrated that 
relaminarisation events are preceded by a weakening of streak instabilities and 
thus necessarily of streamwise vortices, rather than being connected with the streaks themselves.  
Relaminarisation can only occur when the self-sustaining cycle breaks up, and our analysis suggests that this
happens at the streak instability stage.
	}

Concerning the nine-mode model, each data feature has a clear physical interpretation. 
This allows to address
the issue of representativeness versus significance
\citep{jimenez2018machine}. To do so, we suggest to classify the information obtained
into two categories, one comprises features of known phenomenological
importance - the representative features - which can be identified in a shear
flow by the naked eye such as streamwise vortices or streaks, and the other
comprises features of potential dynamical significance that are more difficult
to observe directly, i.e. being not or at least much less representative.  In
the present context, the second class contains modes representing linear streak
instabilities, for instance.  For the first class, SHAP was used to uncover
which of the known representative features were significant for the prediction
of relaminarisation events. First, we see that a known representative feature
of near-wall dynamics, the streamwise vortex, is identified.
%
%
Second, and more interestingly, the dynamics of streak mode, also a
representative feature, is not relevant for the prediction of relaminarisation
events.  In the second class, SHAP identifies a dynamically significant feature
among the streak instabilities, the fundamental spanwise mode that is odd in
$x_2$.  This suggests that even though the streak has several types of
instabilities within the SSP, the main effect on the dynamics with respect to
relaminarisation events stems from spanwise instabilities of odd parity with
respect to reflection about the midplane, at least in the nine-mode model.

For the DNS data, we find that SHAP identifies spatially localised regions 
in the flow that are relevant for the prediction of relaminarisation events. Taking guidance from the 
results obtained for the NMM, a classification task to probe the relevance of 
streak instabilites for the prediction of relaminarisation events was constructed by showing 
2D data planes orthogonal to the spanwise direction to a classifier, that is, planes including cross-sections 
of high- and low-speed streaks. We find that SHAP values cluster in certain regions on the plane connected 
with variations in streak structure, which 
indicates that the classifier bases its decision on proxies for streak modulations corresponding 
to linear streak instabilities within the SSP. \review{Since streamwise vortices are generated by nonlinear
interactions of streak instabilities, SHAP thus identifies the suppression of streak instabilities as the 
point of breakdown within the self-sustaining cycle leading to relaminarisation. }

SHAP thus identifies not only which of the characteristic phenomenological
features of the self-sustaining process in a wall-bounded shear flow are
significant for the prediction of relaminarisation events, it also recognises
patterns in the data corresponding to its fundamental dynamical mechanism. That
is, it serves as a means to distinguish representativeness from significance of
features for a given ML task.  Furthermore, variances in the feature importance
ranking across prediction horizons are commensurate with differences in the
dynamics one would expect closer or further away from a relaminarisation event.

Finally, we conclude with a few suggestions for further work.  As SHAP is
model-agnostic and can be used in conjunction with deep learning algorithms,
this method can be upscaled and applied to high-dimensional experimental and
numerical data. 
Essentially, we can enviseage two main application categories, one aimed at
obtaining further physics insight from high-dimensional numerical or
experimental data, and one at purely technical improvements of analysis
routines.  The former will in most instances require a pre-processing step to
decompose the data into physically interpretable features, while no such
decomposition would be required for the latter to yield useful results.  The
results for the low-dimensional model presented here serve as a proof of 
concept for the former. The aforementioned example calculation using DNS data of transitional minimal 
plane Couette flow, where SHAP was used to identify regions in the flow that the classifier needs to see in order 
to render accurate predictions, demonstrates that the latter is in principle possible. 
An in-depth analysis of DNS data at higher Reynolds number and resolution is beyond the scope of the present 
paper, however it would be a very interesting follow-up study. For the former, examples 
for useful data decompositions are proper orthogonal decomposition
(POD) \citep{Berkooz1993} or dynamic mode decomposition (DMD)
\citep{Schmid2010}, both by now widely used techniques for data analysis and
model reduction in fluid dynamics.  While POD returns modes corresponding to
energy content, DMD decomposes experimental or numerical data into
spatio-temporally coherent structures labelled by frequency. The suitability of
any data decomposition and reduction technique would depend on the planned
task. 

%
Identifying important modes and their interactions for a
data representation without a straightforward physical interpretation 
could be useful to construct for instance a
lower-dimensional description of the dynamics by only retaining important
modes for any given ML task at hand. In complex geometries, SHAP analyses could provide guidance as to which
part of a simulation domain requires high resolution and where compromises
regarding resolution and simulation cost will be less detrimental to the overall
accuracy of the simulation.
This can become particularly helpful outwith pure turbulence research, such as for
active fluids, geo- and astrophysical flows, viscoelastic flows or complex flow
geometries, as these applications are data-intensive and the flow complexity 
often does not allow straightforward feature intepretation.

Ultimately, results reported here are based on a
statistical analysis of the SHAP values. Additionally, as SHAP values can be calculated
alongside real-time predictions, per-sample SHAP values may prove themselves as
useful tools in on-the-fly tasks such as machine-assisted nonlinear flow
control or for optimisation problems.

\vspace{0.5cm}

Declaration of Interests. The authors report no conflict of interest.

\section*{Acknowledgements}

This project originated through discussions with Bruno Eckhardt, who sadly
passed away on August $7^{\rm th}$ 2019. We hope to have continued the work
according to his standards and any shortcomings should be attributed to M.
Linkmann.  The authors thank Tobias Bischoff, Michele Buzzicotti, Peter Veto,
Michael Grau, Xiaojue Zhu, Bernd Noack, Eric Jelli and J\"{o}rg Schumacher for
helpful conversations.  The computations have been carried out on the MaRC2
compute cluster of the Philipps-University of Marburg and on Cirrus
({\tt www.cirrus.ac.uk}) at the University of Edinburgh. We thank the
respective support teams for technical advice. Computing time
on Cirrus was provided through a Scottish Academic Access allocation.
We acknowledge financial support from the German Academic Scholarship Foundation (Studienstiftung des deutschen Volkes) and the Priority Programme SPP
1881 ``Turbulent Superstructures" of the Deutsche Forschungsgemeinschaft (DFG)
under grant Li3694/1.

\appendix

\section{}

\subsection{\label{sec:appendix.boosted_trees}Boosted trees}

In boosted methods, the ensemble of $K$ weak learners, $\{g_k\}_{k=1}^{K}$, is set up in an additive manner, so that the output of the boosted model $g^{(K)}$ for a sample $\vec{z}\in\mathbb{R}^{M}$ is
\begin{equation}
	g^{(K)}(\vec{z}) = \sum_{k=1}^{K} g_k(\vec{z})~\in\mathbb{R}.
\end{equation}
The models $g_k$ are learned sequentially as to correct the mistakes of the previous models $\{g_i\}_{i=1, \dots, k-1}$ without altering them. Given a per-sample loss $L(y, \hat{y})$ between the true sample label $y$ and the prediction of the previous model $\hat{y}=g^{(k-1)}$, the next weak learner $g_k$ is found by optimising
\begin{equation}
	\min_{g_k} \sum_{n=1}^{N_t} L(y_n, g^{(k-1)}(\vec{z}_n)+g_{k}(\vec{z}_n)),
	\label{eq:boosting_objective}
\end{equation}
with $N_t$ as number of training samples.

Boosted trees use decision trees $T(\vec{z}; \vec{\theta})$ \citep{breiman1984classification} as weak learners, $g_k(\vec{z}) = T(\vec{z}; \vec{\theta})$ with $\vec{\theta}$ as parameters of the decision tree. A decision tree classifier is a binary tree that categorises its input as the class of the terminal node the input is assigned to. A decision tree with $J$ terminal nodes evaluates $T(\vec{z}; \vec{\theta})$ according to
\begin{equation}
	T(\vec{z}; \vec{\theta}) = \sum_{j=1}^{J} \gamma_j I(\vec{z}\in R_j)
\end{equation}
with $I$ as indicator function and parameters $\vec{\theta}=\{R_j, \gamma_j\}_{j=1}^{J}$ as terminal regions $R_j$ of the terminal nodes in the input space and the assigned values in the terminal nodes $\gamma_j$.

\begin{figure}
	\centering
	\includegraphics[scale=1]{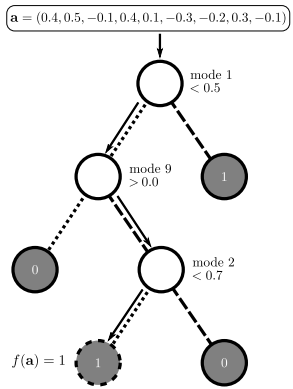}
	\caption{
		Schematic classification of a spectral velocity field $\vec{a}$ by a decision tree. The dotted and dashed lines denote positive and negative decisions, respectively. The $J=3$ terminal notes are coloured in grey and the dashed terminal node marks the output of the example classification.
	}
	\label{fig:machine_learning.explanation.tree_structure}
\end{figure}

Figure~\ref{fig:machine_learning.explanation.tree_structure} illustrates how a spectral velocity field is classified according to an example decision tree with $J=3$. The sample we seek to classify is assigned to the grey terminal node with a dashed border after it transversed the binary tree structure. The structure is made up of binary decisions, which are noted next to the nodes in black. The predicted class of the sample is 1, since the terminal node it has been assigned to is itself assigned the class 1. Learning a tree consists of finding the appropriate tree structure and the terminal node classes in grey.

Boosted trees have been shown to yield state-of-the-art performances on a number of standard classification benchmarks \citep{li2010empirical} and are thereby suitable for the task of classifying relaminarisation events in shear flows.

As finding the optimal tree structure is an intractable problem, gradient boosted trees make use of the gradient of the deployed per-sample loss function $L$ for finding an optimal tree structure $\{R_j\}_{j=1}^J$. The specific role of the gradients in improving the boosted trees depends on the exact model: for example, traditional gradient boosted tree methods \citep{ridgeway2006gbm} fit a decision tree to the negative gradient of $L$, arising from the first order Taylor expansion of equation \eqref{eq:boosting_objective}, in order to benefit from the generalisation ability of the tree to new data \citep[Chapter 10.10]{hastie2009elements}.

The gradient boosted tree algorithm implemented by XGBoost, however, extends the boosting objective \eqref{eq:boosting_objective} about a regularisation term and expands the loss function $\mathcal{L}^{(k)}$ in the next weak learner $g_k$ to be added,
\begin{equation}
	\begin{split}
		\mathcal{L}^{(k)} & = \sum_{n=1}^{N_t} L(y_n, \hat{y}_{n}^{(k-1)}+g_k(\vec{z}_n)) + \Omega(g_k) \\
		& \approx \sum_{n=1}^{N_t} \bigg[L(y_n, \hat{y}_n^{(k-1)})+\frac{\partial L(y_n, d)}{\partial d}\Bigr|_{d=\hat{y}_n^{(k-1)}}g_k(\vec{z}_n) \\
		& \qquad\qquad + \frac{1}{2} \frac{\partial^2 L(y_n, d)}{\partial d^2}\Bigr|_{d=\hat{y}_n^{(k-1)}} g_k^2(\vec{z}_n) \bigg] + \Omega(g_k)\\
	\end{split}
	\label{eq:iterative_loss_function}
\end{equation}
as iterative objective for iteration step $k$ with regularisation $\Omega(g_k)=\omega J_k + \frac{1}{2} \lambda |(\gamma_{1k}, \dots, \gamma_{J_k k})|^2$ on tree $k$. Here, $J_k$ denotes the number of terminal nodes of tree $k$ and $\gamma_{jk}$ with $j=1,\dots,J_k$ is the terminal node weight of index $j$ and tree $k$.
Using $R_{jk}$ as terminal node set of index $j$ and tree $k$, the objective is used to quantify the quality of a tree structure after minimising the quadratic iterative objective,
\begin{equation}
	\mathcal{L}^{(k)}(\{R_{jk}\}_{j=1,\dots,J_k}) = -\frac{1}{2}\sum_{j=1}^{J_k}\frac{(\sum_{i\in I_j}h_i^{(1)})^2}{\sum_{i\in I_j}h_i^{(2)} + \lambda}+\omega J_k
\end{equation}
with $I_j=\{i|q(\vec{z}_i)=j\}$ as terminal node index set, $q(\vec{z})$ mapping a sample to the index of the terminal node it is assigned to and $h^{(1)}$ and $h^{(2)}$ as first and second order derivatives of $L$ from Eq.~\eqref{eq:iterative_loss_function}, respectively. Using $\mathcal{L}^{(k)}(\{R_{jk}\}_{j=1,\dots,J_k})$, the quality of a node $I$ to be split into left and right, $I=I_L\cup I_R$, can be measured quantitatively.

The remarkable novelty of XGBoost, aside from the regularisation $\Omega(g_k)$, is a split finding technique that uses weighted data quantiles to find appropriate split candidates, which are themselves evaluated with a novel approximate split finding algorithm that uses the split loss $\mathcal{L}^{(k)}(\{R_{jk}\}_{j=1,\dots,J})$ as metric. Furthermore, XGBoost scales to very large datasets as all components are properly parallelisable. This is due to additional technical innovations, such as data sparsity awareness, cache awareness and out-of-core computations if the dataset gets very large.

The loss function $L$ is the logistic regression for binary classification with output probabilities,
\begin{equation}
	L(y, \hat{y}) = y \log(\sigma(\hat{y})) + (1-y) \log(1-\sigma(\hat{y})),
\end{equation}
with logistic function $\sigma(y)=1/(1+\exp(-y))$.

XGBoost classifiers come with a number of tunable hyperparameter that need to be specified by the user. The following parameters are tuned in this work and are therefore explained here using the XGBoost notation:

\begin{align*}
	\text{ \emphasize{n\_estimators} } (h_{NE}) \text{ - } & \text{ The number of decision trees to fit to the task.} \\
	& \text{ This number corresponds to $K$ in the notation } \\
	& \text{ introduced above.} \\
	\text{ \emphasize{max\_depth} } (h_{MD}) \text{ - } & \text{ The maximal depth of each decision tree.} \\
	\text{ \emphasize{min\_child\_weight} } (h_{MCW}) \text{ - }  & \text{ Regularisation of the tree building process as } \\
	& \text{ to only split a terminal node when it contains } \\
	& \text{ samples of different target classes. } \\
	\text{ \emphasize{gamma} } (h_{GA}) \text{ - } & \text{ Regularisation of the tree building process by } \\
	& \text{ only splitting a terminal node when the loss } \\
	& \text{ reduces at least about a value of \emphasize{gamma}. } \\
	\text{ \emphasize{subsample} } (h_{SS}) \text{ - } & \text{ The percentage of the number of overall training } \\
	& \text{ samples $N_t$ that are randomly sampled per decision tree.} \\
	\text{ \emphasize{colsample\_bytree} } (h_{CBT}) \text{ - } & \text{ The percentage of the number of training } \\
	& \text{ features $M$ that are randomly sampled per decision tree.} \\
	\text{ \emphasize{learning\_rate} } (h_{LR}) \text{ - } & \text{ Scaling applied to the learned terminal node } \\
	& \text{ weights $\gamma_{jk}$ after each boosting iteration \citep{friedman2002stochastic}, } \\
	& g_k(\vec{z}) = g_{k-1}(\vec{z}) + \sum_{j=1}^{J_k}h_{LR}\text{ } \gamma_{jk} I(\vec{z}\in R_{jk}).
\end{align*}

\subsection{\label{sec:appendix.optimal_hyperparameters}Optimal XGBoost hyperparameters for relaminarisation prediction}

Table~\ref{tab:appendix.optimal_hyperparamters} lists the optimal hyperparameters as identified by the randomised hyperparameter optimisation with $100$ drawn hyperparameter samples. Since all prediction times use the same $100$ hyperparameter samples, the same set of hyperparameters might be found to be optimal for more than one prediction horizon, e.g. $t_{p}=200$ and $250$.

\begin{table*}
	\begin{center}
	\begin{tabular}{lrcccccc}
		\hline
        $t_{p}$ & $h_{NE}$ & $h_{MD}$ & $h_{MCW}$ & $h_{GA}$ & $h_{SS}$ & $h_{CBT}$ & $h_{LR}$ \\ \hline
        $200$ &  $941$ & $42$ & $5$ & $0.177$ & $0.572$ & $0.823$ & $0.173$  \\
        $250$ &  $941$ & $42$ & $5$ & $0.177$ & $0.572$ & $0.823$ & $0.173$  \\
        $300$ &  $511$ & $44$ & $1$ & $1.302$ & $0.960$ & $0.992$ & $0.215$  \\
        $350$ &  $765$ & $33$ & $2$ & $3.903$ & $0.769$ & $0.731$ & $0.048$  \\
        $400$ & $1148$ & $39$ & $8$ & $0.501$ & $0.917$ & $0.620$ & $0.008$ \\
        $450$ & $1148$ & $39$ & $8$ & $0.501$ & $0.917$ & $0.620$ & $0.008$ \\ \hline
	\end{tabular}
	\caption{Table of optimal hyperparameters for XGBoost classifier for the task of predicting the relaminarisation of the turbulent trajectory. The abbreviations in the header line have been introduced in the main text of Appendix~\ref{sec:appendix.boosted_trees}.}
	\label{tab:appendix.optimal_hyperparamters}
	\end{center}
\end{table*}

\newpage
\bibliographystyle{jfm}
\bibliography{bibliography}

\end{document}